\documentclass{emulateapj}
\usepackage{apjfonts}
\usepackage{amsmath}
\usepackage{lscape}
\usepackage{mathrsfs}
\usepackage{epsfig}
\usepackage{color}
\usepackage{textcomp}
\slugcomment{Appears in the Astrophysical Journal, 2010, 710, 878-885}

\def\calcb{{\cal C}_{B}}

\def\mh{M_{\rm H}}

\def\rd{{\rm d}}

\def\sunm{M_{\odot}}

\def\ergs{\ifmmode {\rm ergs~ s^{-1}} \else {\rm ergs~s^{-1}}\ \fi}
\def\kms{\ifmmode {\rm km~ s^{-1}} \else {\rm km~s^{-1}}\ \fi}

\def\mbh{M_{\bullet}}

\def\mgii{\ifmmode Mg {\sc ii} \else Mg {\sc ii}\ \fi}

\def\sunm{M_{\odot}}

\def\lax{{$\mathrel{\hbox{\rlap{\hbox{\lower4pt\hbox{$\sim$}}}\hbox{$<$}}}$}}
\def\gax{{$\mathrel{\hbox{\rlap{\hbox{\lower4pt\hbox{$\sim$}}}\hbox{$>$}}}$}}

\def\ergs{${\rm erg~s^{-1}}$}

\begin{document}

\title{Episodic activities of supermassive black holes at redshift $z\le 2$:
driven by mergers?}

\author{
Yan-Rong Li\altaffilmark{1,3},
Jian-Min Wang\altaffilmark{1,4},
Ye-Fei Yuan\altaffilmark{2},
Chen Hu\altaffilmark{1}, and
Shu Zhang\altaffilmark{1}
}
\affil{$^1$ Key Laboratory for Particle Astrophysics, Institute of High Energy Physics, CAS, 
19B Yuquan Road, Beijing 100049, China\\
$^2$ Center for Astrophysics, University of Science and Technology of China, Hefei
230026, China\\
$^3$ Graduate University of Chinese Academy of Sciences, 19A Yuquan Road, Beijing 100049,
China \\
$^4$ Theoretical Physics Center for Science Facilities, Chinese Academy of Sciences,
China}

\begin{abstract}
It has been suggested for quite a long time that galaxy mergers trigger activities
of supermassive black holes (SMBHs) on the grounds of imaging observations of 
individual galaxies. To quantitatively examine this hypothesis,
we calculate quasar luminosity functions (LFs) by manipulating the observed galaxy LFs ($z\lesssim 2$) 
and theoretical merger rates from semi-analytical formulations. We find that the model 
reproduces the observed quasar LFs provided that the mass ratio ($q$) of the secondary galaxy to 
the newly formed one changes with cosmic time. The results show that the fraction of major mergers 
decreases from $f_{\rm maj}\sim 0.2$ at $z\sim 2$ to $f_{\rm maj}\rightarrow 0$ at $z\sim 0$. 
As a consequence, the newly formed SMBHs from major mergers at $z\sim 2$ may acquire 
a maximal spin due to the orbital angular momentum of the merging holes. Subsequently, 
random accretion led by minor mergers rapidly drives the SMBHs  to spin down. 
Such an evolutionary trend of the SMBH spins is consistent with that radiative
efficiency of accreting SMBHs strongly declines with cosmic time, reported by Wang et al. (2009). 
This suggests that minor mergers are important
in triggering activities of SMBHs at low redshift while major mergers may dominate at high redshift.

\end{abstract}

\section{Introduction}
In the hierarchical model of galaxy formation, quasar activities are related to the mergers of
galaxies (e.g., Toomre 1977; Hernquist 1989; Barnes \& Hernquist 1992; Hopkins et al. 2008). 
During the merging processes, tidal torque drives gas inflows into the inner region of galaxies,
forming starbursts and igniting the central SMBHs (e.g., Hernquist 1989; Barnes \& Hernquist 1996; 
Springel et al. 2005). Once quasar activities are triggered,
the mutual interaction between the SMBHs and galaxies is unavoidable accompanied by feedback  
of active galactic nuclei (AGNs) (e.g., outflows or radiation; see Di Matteo et al. 2005; Croton et al. 2006; 
Bower et al. 2006). This is believed to regulate coevolution of SMBHs and their host spheroids, 
inferred from the well-established correlation between
SMBH mass and spheroid mass or dispersion velocity (Magorrian et al. 1998; Gebhardt et al.
2000; Ferrarese \& Merritt 2000; Marconi \& Hunt 2003; H\"aring \& Rix 2004), 
however, the details remain open.

\begin{deluxetable*}{lccccc} 
\tabletypesize{\footnotesize}
\tablecolumns{6} 
\tablewidth{\textwidth} 
\tablecaption{Model list}
\tablehead{ 
\colhead{Reference}     &\colhead{Model}             &\colhead{Merger type\tablenotemark{a}} & \colhead{Quasar light curve\tablenotemark{b}}&\multicolumn{2}{c}{$\mbh-\mh$ relation\tablenotemark{c}}\\\cline{5-6}
\colhead{} &\colhead{}&\colhead{}&\colhead{}  &\colhead{ $z$-dependent} &\colhead{ $\mh$-dependent}
} 
\startdata 
R85       & Star disruption + galaxy merger rate      & Yes           &\nodata 
          &\nodata                  &\nodata\\
C90       &  Halo merger rate                         & Major         &Gau        
          &\nodata                  &\nodata\\
HR93      & Halo formation rate                     & \nodata       &Exp            
          &$\surd$                  &\texttimes\\  
HL98      & Halo formation rate                     & \nodata       &Exp           
          &\texttimes               &\texttimes\\
WL02      & Halo mass function + halo merger rate   & Yes           &Step           
          &$\surd$                  &$\surd$\\
WL03      & Halo mass function + halo merger rate   & Major         &Step           
          &$\surd$                  &$\surd$\\
VHM03     & Merger tree                             & Major         &Step        
          &\texttimes               &$\surd$\\
HHCK08    &Halo mass function + galaxy merger rate  & Major         & Pow           
          &$\surd$                  &$\surd$\\
S09       & $N$-body merger rate                    & Major         & Exp + Pow     
          &$\surd$                  &$\surd$\\
This work &Galaxy LFs + halo merger rate           &$z$-dependent   &Step           
          &$\surd$                  &~~~~$\surd$
\enddata
\tablerefs{R85: Roos (1985); C90: Carlberg (1990); HR93: Haehnelt \& Rees (1993);
HL98: Haiman \& Loeb (1998); WL02: Wyithe \& Loeb (2002); WL03: Wyithe \& Loeb (2003);
VHM03: Volonteri et al. (2003); HHCK08: Hopkins et al. (2008); S09: Shen (2009).}
\tablenotetext{a}{ ``Yes'' means no specifying merger type.}
\tablenotetext{b}{ ``Exp'' means exponential, ``Pow'' means pow law, ``Step'' means step function, and
``Gau'' means the Gaussian distribution of quasar lifetime..}
\tablenotetext{c}{``$\surd$'' means dependent and ``\texttimes'' means independent.}
\end{deluxetable*}

Based on the long-standing ``merger hypothesis'' (Toomre 1977),
quasar LFs have been extensively modeled within the framework of hierarchical structure
formation (Roos 1985; Carlberg 1990; Haehnelt \& Rees 1993; Haiman \& Loeb 1998; Wyithe \& Loeb 2002, 2003; 
Volonteri et al. 2003; Hopkins et al. 2008; Shen 2009). A brief summary of hierarchical models
from the published literature is
given in Table 1. We note that most models employ major mergers.
In these models, the scaling relations between black hole mass and dark matter halo mass
 are generally employed to combine the mergers of galaxies with dark matter halos. The dark matter halos follow
the extended Press-Schechter (EPS) theory  (e.g., Lacey \& Cole 1993). After a merger event,
a quasar is ignited and shines at the Eddington luminosity with a universal light curve.
The rapid decreases as a main feature of quasar LFs can be well produced by the 
decrease in the merger rates and the availability of cold gas to fuel the SMBHs after 
the peak activities of SMBHs. Although these models quantitatively describe the shape of quasar LFs, 
it is still insufficient to understand the cosmological evolution of quasars in light of merger hypothesis.

There is increasing evidence showing that major mergers cannot constitute the only triggering mechanism of
quasar activities from observations based on the studies of absorption properties of 
AGNs (e.g., Hasinger 2008) or the galaxy morphologies (e.g., Georgakakis et al. 2009; Reichard et al. 2009), 
as well as from theoretical work by modeling LFs of low-level AGNs (e.g., Hopkins \& Hernquist 2006; 
Marulli et al. 2007). It is evident that minor mergers between 
galaxies and smaller satellites occur much more frequently (Barnes \& Hernquist 1992; Mihos \& Hernquist 1994).
Many previous works have investigated the role of minor mergers in triggering quasar activities 
(Gaskell 1985; Hernquist 1989; Hernquist \& Mihos 1995; De Robertis et al. 1998; Taniguchi \& Wada 1996; 
Taniguchi 1999; Corbin 2000; Chatzichristou 2002; Kendall et al. 2003; Hopkins \& Hernquist 2009a).
Assuming that growth of SMBH occurs via episodic accretion in light of Soltan's argument (Soltan 1982), 
Wang et al. (2009, hereafter W09) derived an $\eta$-equation 
to describe cosmological evolution of radiative efficiency of SMBHs.
It is found that the radiative efficiency does exhibit
a strong cosmological evolution: it decreases from $\eta\sim 0.3$ at $z\sim 2$ to $\eta\sim 0.03$ at $z\sim 0$.
This can be interpreted by the mounting importance of minor mergers towards low redshifts. 
At high redshift, major mergers produce fast rotating SMBHs due to conservation of (orbital) angular momentum of the merging holes. Subsequently random accretion, which may be related to
minor mergers (Kendall et al. 2003; King \& Pringle 2007), acts to spin them down efficiently. 

In this paper, we extend previous model (Wyithe \& Loeb 2002) and construct quasar LFs from the observed
galaxy LFs and theoretical merger rates. Our goal is to investigate the redshift-dependent merger history and the
resulting spin evolution of SMBHs. We describe our model in
Section 2. In Section 3, we present comparisons of the modeled quasar LFs with the observational data, followed by
discussions on the merger history and accretion history, and the spin evolution of SMBHs. 
The conclusions are drawn in Section 4. 
We use cosmological constants $\Omega_m=0.3$, $\Omega_{\Lambda}=0.7$, the Hubble constant 
$H_0=70~{\rm km~s^{-1}~Mpc^{-1}}$ and $\sigma_8=0.87$ throughout the paper.

\figurenum{1}
\begin{figure*}[ht!]
\centering
\includegraphics[angle=-90,width=0.6\textwidth]{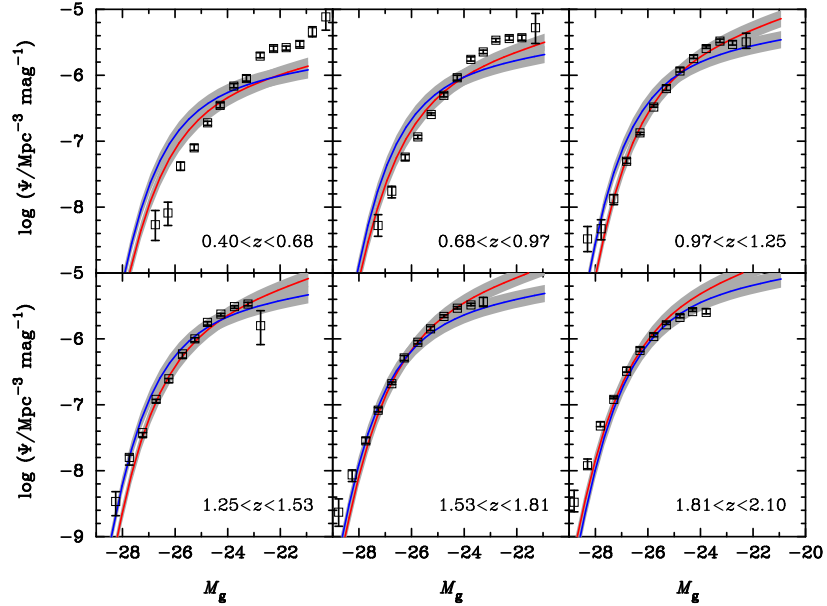}
\figcaption{\footnotesize Comparisons of the predicted luminosity functions
with observations. The values of the parameters are given in Table 2. 
Squares show the observation data from Croom et al. (2009). 
Red and blue lines are modeled by Ilbert et al. (2005) and Dahlen et al.
(2005), respectively. Shared areas represent the Poisson statistical
uncertainties $\Delta \Psi/\Psi=0.3$ from the uncertainties of galaxy LFs. The conversion 
$M_{\rm g}(z=2)=M_B(z=0)-0.527$ 
(Croom et al. 2009) is used. }
\label{fig1}
\end{figure*}
\begin{deluxetable*}{lccc} 
\tabletypesize{\footnotesize}
\tablecolumns{4} 
\tablewidth{0.8\textwidth} 
\tablecaption{The free parameters in the models}
\tablehead{ 
\colhead{Parameter} &\multicolumn{2}{c}{Given by galaxy LFs of \tablenotemark{a}} & \colhead{Implications}\\\cline{2-3}
\colhead{} & \colhead{Ilbert et al. (2005)} & \colhead{Dahlen et al. (2005)} &\colhead{}
} 
\startdata 
$\log(\epsilon_0)$ & $-3.3\pm0.7$     &$-5.0\pm0.6$&The initial ratio of halo to SMBH mass\\
$\log(t_{\rm dc,0}/{\rm yr})$& $7.54\pm0.04$    &$7.30\pm0.04$   &Characterized timescale of a single episode\\
$\lambda$    & $0.10\pm0.01$        &$0.20\pm0.01$   &The mean Eddington ratio of quasars \\
$q(z)$       & \nodata      &\nodata  &$\Delta \mbh/\mbh$ as a function of $z$
\enddata
\tablecomments{The $\chi_{\rm min}^2/{\rm d.o.f.}$ of fittings are 78/65 and 121/65 for Ilbert et al. 
and Dahlen et al., respectively.}
\tablenotetext{a}{Errors are calculated by assuming $\Delta\Psi/\Psi=0.1$.}
\end{deluxetable*}
\section{Mergers and black hole activities}
\subsection{Merger rates of galaxies}
It is generally assumed that galaxies reside in the central region of dark matter halos, and consequently 
galaxy mergers follow dark matter halo mergers. In our calculations, this is implemented through 
the relationship between the masses of central SMBHs in galaxies and their host dark matter halos. 
First, we present merger rates of dark matter halos based on the EPS theory in this section. 
The probability for a halo of mass $M_1$ at time $t$ merging with another halo of mass $\Delta M$
is given by (Lacey \& Cole 1993)
\begin{eqnarray}
{\dot{\cal R}}=&&\nonumber
  \left(\frac{2}{\pi}\right)^{\frac{1}{2}}\frac{1}{t}\frac{1}{M_2}
  \left|\frac{d\ln \delta_c}{d\ln t}\right|
  \left|\frac{d\ln \sigma_2}{d\ln M_2}\right|
  \frac{\delta_c}{\sigma_2}\left(1-\frac{\sigma_2^2}{\sigma_1^2}\right)^{-\frac{3}{2}}\\
  &&\times\exp\left[-\frac{\delta_c^2}{2}\left(\frac{1}{\sigma_2^2}-
  \frac{1}{\sigma_1^2}\right)\right],
\end{eqnarray}
where $M_2=M_1+\Delta M$ is the mass of newly formed halo, $\sigma_1=\sigma(M_1)$, and $\sigma_2=\sigma(M_2)$. 
The expression of $\sigma$ is given by
\begin{equation}
\sigma^2(M)=\int_0^{\infty}\frac{dk}{2\pi^2}k^2P(k)\left[\frac{3j(kR)}{kR}\right]^2,
\end{equation}
where $j(x)=(\sin x-x\cos x)/x^2$, $\rho_m=3H^2\Omega_m/8\pi G$, and $M=4\pi \rho_mR^3/3$ is the mass within a comoving 
sphere of radius $R$, $G$ is the gravity constant, $H$ is the Hubble constant, and
$\Omega_m$ is the fraction of baryons. The primordial power-law spectrum is
\begin{equation}
P(k)=p_0 kT^2(k),
\end{equation}
where the transfer function is
$T(k)=\left(1+\sum_{i=1}^4c_iq^i\right)^{-1/4}$ $\left(c_5q\right)^{-1}\ln(1+c_5q)$,
$c_1=3.89, c_2=259.21, c_3=162.77, c_4=2027.17, c_5=2.34$, $p_0$ is determined by
observations of $\sigma_8$ at $8h^{-1}~{\rm Mpc}$ and $q=k/(\Omega_mh^2{\rm Mpc^{-1}})$,
where $h$ is the Hubble constant in a unit of 100 km s$^{-1}$ Mpc$^{-1}$.  
$\delta_c(t)$ is expressed by
\begin{equation}
\delta_c(t)=\delta_{c0}\left(\frac{t_0}{t}\right)^{2/3}
           =1.69\left(\frac{t_0}{t}\right)^{2/3},
\end{equation}
where $t_0$ is a referenced epoch.
\figurenum{2}
\begin{figure}
\centering
\includegraphics[angle=-90,width=0.4\textwidth]{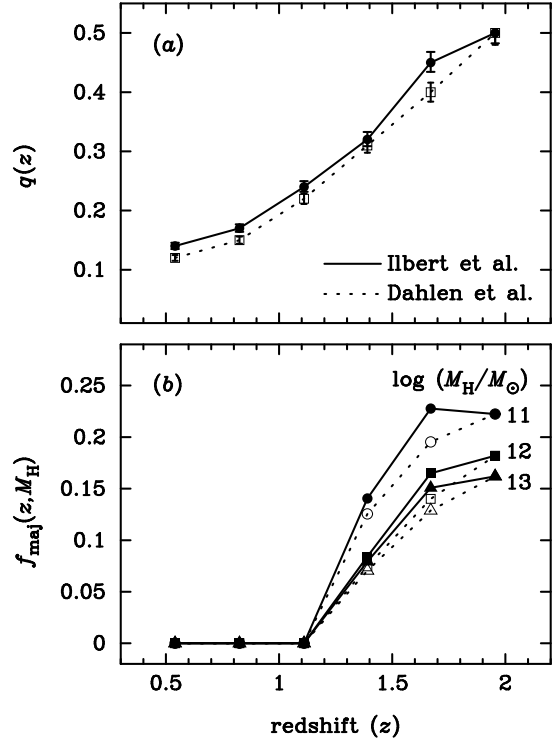}
\figcaption{\footnotesize ({\em a}) Cosmological changes of the mass ratio ($q$), and
({\em b}) the fraction of major mergers to total ($f_{\rm maj}$) as a function of redshift, 
indicating the increasing role of major mergers with redshift. Solid and dashed lines (with symbols) are for
Ilbert et al. (2005) and Dahlen et al. (2005), respectively. Error bars are calculated by assuming 
$\Delta\Psi/\Psi=0.1.$}
\label{fig2}
\end{figure}
\begin{figure}
\figurenum{3}
\centering
\includegraphics[angle=-90,width=0.4\textwidth]{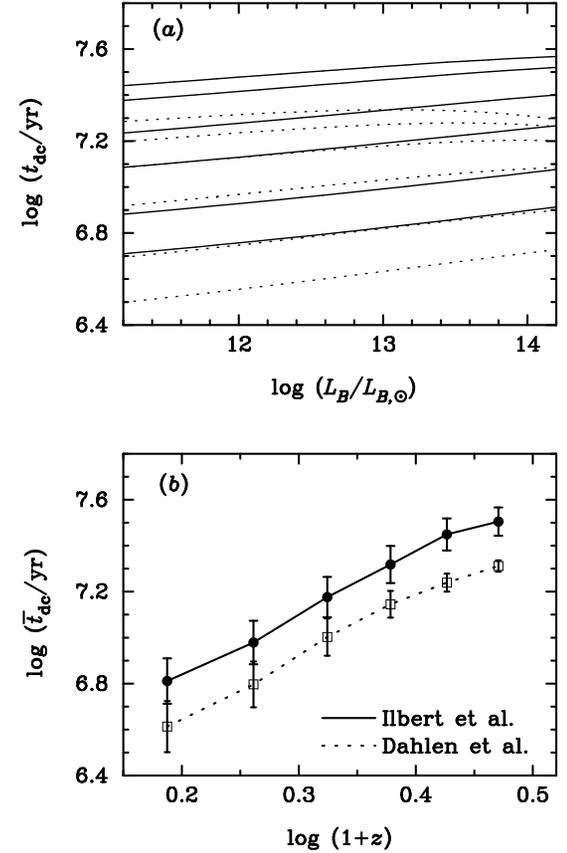}
\figcaption{\footnotesize ({\em a}) Duty cycle time of black hole activity as a function of
quasar luminosity at the redshift bins corresponding to Figure 1 from bottom to top. 
({\em b}) Luminosity-averaged duty cycle time with redshift. Error bars are the
upper and lower limits of the duty cycle time in ({\em a}). Solid and dashed lines (with symbols) are for
Ilbert et al. (2005) and Dahlen et al. (2005), respectively. }
\label{fig3}
\end{figure}
\subsection{Quasar luminosity functions}
In modeling quasar LFs, we extend the method of Wyithe \& Loeb (2002) 
by employing the realistic galaxy LFs
instead of the halo mass function and considering the redshift-dependent merger history.
To derive quasar LFs from {\em known} galaxy LFs through mergers, we assume
that (1) each galaxy hosts an SMBH; (2) all mergers trigger SMBH activities;
(3) a merger simultaneously triggers an SMBH
activity. We neglect the time delay between a galaxy merger and igniting of the central SMBH. 

We use the scaling relation between the black hole mass and the dispersion velocity
$\mbh\propto \sigma^4$ (Tremaine et al. 2002) to connect SMBHs with dark matter halos.
The dispersion velocity correlates with the Keplerian velocity of the dark matter halo via
$v_c\propto \sigma^{\beta}$  with $\beta\approx 0.84$ (Ferrarese 2002), where $v_c$ is
determined by $\mh$ (Barkana \& Loeb 2001)
\begin{equation}
v_c=159.4M_{\rm H,12}^{1/3}\left(\frac{\Omega_m}{\Omega_{m,z}}
    \frac{\Delta_c}{18\pi^2}\right)^{1/6}(1+z)^{1/2}~{\rm km~s^{-1}},
\end{equation}
where $M_{\rm H,12}=\mh/10^{12}h\sunm$ is the dark matter halo mass,
$\Delta_c=18\pi^2+82d-39d^2$ is the final over-density
relative to the critical density at redshift $z$, $d=\Omega_{m,z}-1$ and
$\Omega_{m,z}=\left[\Omega_m(1+z)^3\right]/\left[\Omega_m(1+z)^3+\Omega_{\Lambda}\right]$.
The ratio of SMBH mass to halo mass is then computed as (Wyithe \& Loeb 2002)
\begin{equation}
\epsilon=\frac{\mbh}{\mh}=\epsilon_0M_{\rm H,12}^{\gamma/3-1}
         \left(\frac{\Omega_m}{\Omega_m^z}\frac{\Delta_c}{18\pi^2}\right)^{1/6}
         h^{-1}(1+z)^{\gamma/2},
\end{equation}
where $\gamma=4/\beta=4.76$ and $\epsilon_0$ is a free parameter.
We note that $\epsilon$ is weakly dependent on $\mh$, but sensitive to redshift.
The nonlinear relation between SMBH mass and halo mass with $\gamma>3$ is due to 
the significant baryon accretion during SMBH activities visible to observers (Yu \& Tremaine 2002).
It is worth pointing out that the potential evolution of $\gamma$ is poorly understood
(e.g., Shields et al.
2003; Jahnke et al. 2009; Shields et al. 2006; Alexander et al. 2008; Coppin et al.
2008). In our model, we take
$\gamma=4.76$ and neglect the potential effects of its evolution. 
The modeled quasar LFs are found to be insensitive to $\gamma$.

Once a quasar is triggered, it is assumed to shine with 
a universal $B-$band light curve as (see also Haiman \& Loeb 1998)
\begin{equation}
L_B(t)=\mbh f(t)=\epsilon \mh f(t), ~~~(\mh>M_{\rm H,min})
\end{equation}
where $f(t)$ is a function of time 
and $M_{\rm H,min}$ is the minimum mass of the halo, inside which 
an SMBH can form. Here we simply use a step function for $f(t)$ (Wyithe \& Loeb 2002),
\begin{equation}\label{licu}
f(t)= \frac{\lambda L_{{\rm Edd}}}{\calcb M_{\bullet}}\Theta
      \left(\frac{\Delta \mh}{\mh}t_{\rm dc,0}-t\right),
\end{equation}
where $L_{\rm Edd}=1.38\times10^{38}(\mbh/M_\odot)$ is the Eddington luminosity,
$\lambda$ is the Eddington ratio, $\Delta \mh$ is the mass of the secondary halo, 
$\Theta(t)$ is the Heaviside step function, 
$\calcb$ is the bolometric correction factor defined as the bolometric luminosity 
$L_{\rm Bol}=\calcb L_B$ (Marconi et al. 2004). 
This light curve takes into account that the gas deposited from the satellite galaxy
runs out in a time proportional to $\Delta\mh/\mh$ 
\footnote{Basically, the gas from the satellite galaxy is
\begin{equation}\nonumber
\Delta M_{\rm gas}=\frac{\Delta M_{\rm gas}}{\Delta\mh}\frac{\Delta\mh}{\mh}\mh
\propto \frac{\Delta\mh}{\mh}\mbh,
\end{equation}
where the ratios $\epsilon=\mbh/\mh$ and $\Delta M_{\rm gas}/\Delta\mh$ are assumed as constants 
in a zeroth-order approximation. As accretion of the SMBH is limited at the Eddington rate, 
the gas will run out via accretion in a time proportional to $(\Delta\mh/\mh)$.
} 
(Wyithe \& Loeb 2002).
It turns out that spectral energy distributions (SEDs) of AGNs have no 
appreciable evolution with cosmic time (e.g., Shemmer et al. 2008). This would guarantee that Eddington 
ratio ($\lambda$) may also be independent of redshift, according to its correlation with the SEDs 
(Wang et al. 2004). A similar conclusion has been drawn through an analysis of a large quasar sample 
from the Sloan Digital Sky Survey (SDSS; Shen et al. 2008) and the AGN and Galaxy Evolution Survey 
(AGES; Kollmeier et al. 2006), respectively.
Therefore, we assume that the triggered SMBHs shine at a fixed Eddington ratio. 
The value of $\lambda$ can be obtained from fittings of quasar LFs.

With these assumptions, the modeled quasar LFs can be written as
(Wyithe \& Loeb 2002)
\begin{eqnarray}
\Psi(L_B,z)=&&\nonumber
                  \int_z^{\infty}\rd z'\int_{M_{\bullet,{\rm min}}}^{\infty}\rd\mbh
                  \int_{0}^{q\mbh}\delta\left[L_B-\mbh f(t_z-t')\right]\\
               &&\times \left|\frac{\rd^2N_{\rm merg}}{\rd\Delta \mbh \rd t}\right|_{\mbh'}
                  \left|\frac{\rd n_{\bullet}}{\rd\mbh}\right|_{\mbh^{\prime}}
                  \frac{\rd t'}{\rd z'}\rd\Delta \mbh,
\end{eqnarray}
where $t_z$ and $t'$ are the cosmic time at redshift $z$ and $z'$, $\delta(x)$ is the Dirac delta function,
$\left|\rd n_{\bullet}/\rd\mbh\right|_{\mbh^{\prime}}$ is the number density
of black holes,
$\left|\left(\rd^2N_{\rm merg}\right)/\left(\rd\Delta \mbh \rd t\right)\right|_{\mbh'}$
is the merger rate, $\mbh^{\prime}=\mbh-\Delta \mbh$ and the mass ratio
\begin{equation}
q(z)=\frac{\Delta \mbh}{\mbh}=\frac{\mbh-\mbh^{\prime}}{\mbh}.
\end{equation}
Note that a major merger is defined as $0.5\geq q\geq0.25$ so that the mass ratio of the two merging galaxies 
is larger than 1:3. The upper limit forces the secondary galaxy to always be smaller than the primary one.
We treat $q$ as a function of redshift and constrain it from the fittings of quasar LFs.

Integrating over $\mbh$, we can find
\begin{eqnarray}
\Psi(L_B,z)=&&\nonumber
                  \int_z^{\infty}\rd z' \frac{L_B}{\mbh f(t_z-t')}
                  \int_{0}^{q\mbh}
                 \left|\frac{\rd^2N_{\rm merg}}{\rd\Delta \mbh \rd t}\right|_{\mbh'}\\
                &&\times  \left|\frac{\rd n_{\bullet}}{\rd\mbh}\right|_{\mbh^{\prime}}
                  \frac{\rd t'}{\rd z'}\rd\Delta \mbh.
\end{eqnarray}
Quasars in the cosmic time range ($t_z-t_Q$, $t_z$) with a lifetime $t_Q=(\Delta \mh/\mh)t_{\rm dc,0}$ 
contribute to the LFs at cosmic time $t_z$.
Since the lifetime is much smaller than the cosmic time $t_z\sim H(z)^{-1}$ at the redshift $z<2$, 
we can then integrate over $z'$ by replacing $dt'$ with $(\Delta \mh/\mh)t_{\rm dc,0}$.
We employ the galaxy LFs, defined as $\Phi(L',z)$, to trace the SMBH mass function $\rd n_{\bullet}/\rd\mbh$, and 
use the ratio $(\mbh/M_{\rm H})$ to calculate the merger rate of SMBHs from that of dark matter halos,
\begin{equation}
\left|\frac{\rd^2N_{\rm merg}}{\rd\Delta \mbh \rd t}\right|_{\mbh^{\prime}}
     =\frac{3}{\gamma\epsilon}\left|\frac{\rd^2N_{\rm merg}}{\rd\Delta \mh \rd t}\right|_{\mh'}
     =\frac{3}{\gamma\epsilon}\dot{{\cal R}},
\end{equation}
where $\mh'=\mh-\Delta \mh$. Finally we get the quasar LFs

\begin{equation}\label{lfeqn}
\Psi(L_B,z)=t_{\rm dc,0}\frac{\rd L'}{\rd L_B}\int_{0.05\mh}^{q\mh}\dot{{\cal R}}\Phi(L',z) 
\frac{\Delta \mh}{\mh}\rd \Delta M_{\rm H},
\end{equation}
where the factor $(\rd L'/\rd L_B)$ comes from the different luminosity bands used in the galaxy and quasar LFs, 
and $L'$ is the luminosity of galaxies hosted in the halos with a mass $\mh-\Delta\mh$. 
We calculate $L'$ using the relation between the $R-$magnitude of normal galaxies and the mass of black holes  
from McLure \& Dunlop (2002; see the appendix). Note that the low limit in Equation (13) is set so that the
secondary galaxy should be massive enough to provide significant gas for the mass assembly of 
the newly formed black hole. 

With the merger rates, we have a duty cycle time
\footnote{Note that here ``duty cycle time'' refers to 
the active time of the quasars in a timescale $H_z^{-1}$.}
of the quasars during a Hubble time $H_z^{-1}$
\begin{equation}\label{duti}
t_{\rm dc}(L_B,z)=t_{\rm dc,0}\int_{0.05\mh}^{q\mh} \frac{\Delta\mh}{\mh}H_z^{-1}\dot{{\cal R}}\rd \Delta M_{\rm H},
\end{equation}
where $H_z$ is the Hubble constant at redshift $z$. We also define a fraction of major mergers to the total as
\begin{equation}
f_{\rm maj}(z,\mh)=\frac{\dot{N}_{\rm maj}(z,\mh)}{\dot{N}_{\rm tot}(z,\mh)},
\end{equation}
where 
\begin{equation}
\dot{N}_{\rm maj}(z,\mh)=\int_{0.25\mh}^{q\mh}\dot{\cal R}(z)\rd \Delta \mh,
\end{equation}
and
\begin{equation}
\dot{N}_{\rm tot}(z,\mh)=\int_{0.05\mh}^{q\mh}\dot{\cal R}(z)\rd \Delta \mh.
\end{equation} 
According to the definition of a major merger, when $q(z)\le 0.25$, we have $f_{\rm maj}=0$. 
We stress here that the parameter $f_{\rm maj}(z)$ conveniently traces the gradual evolution from
major to minor mergers, showing a merger history.

As a summary, Table 2 gives a list of the free parameters used in our
model. The parameter $q$ determines both the shape and the number density of quasars.
The amplitude of quasar LFs is linearly proportional to $t_{\rm dc,0}$, which is easily
justified from Equation (13).
The other parameters ($\epsilon_0$ and $\lambda$) can be independently quantified by
theoretical or observational arguments. Fitting quasar LFs would provide more precise
values of these parameters deliberately. 

\section{Comparison with observations}
\subsection{Modeling quasar LFs}
We use two sets of galaxy LFs well-determined by {\em Hubble Space Telescope} and the Great Observatories Origins Deep Survey (GOODS; Dahlen et al. 2005), and VIMOS-VLT Deep
Survey (VVDS; Ilbert et al. 2005), respectively. 
Since both sets of galaxy LFs are only available for $z\lesssim 2.0$, the modeled quasar LFs are then
also limited. Galaxy LFs are generally described by,
\begin{equation}
\Phi(L,z)=\Phi_* \left(\frac{L}{L_*}\right)^{\alpha}\exp\left({-\frac{L}{L_*}}\right),
\end{equation}
where $\Phi_*$ is a constant, $\alpha$ is an index and $L_*$ is the cutoff luminosity of
galaxies. We have to keep in mind that the galaxy LFs are based on photometric
redshifts. We use quasar LFs determined by the combined data of the Two-Degree Field survey
and the SDSS, which to date are the best constrained
LFs with unprecedented precision (10,637 quasars) and the dynamic range (g-band flux limit of 21.85;
Croom et al. 2009).

Quasar LFs are characterized by double power-laws with a break luminosity (e.g., Hopkins et al. 2007).
In our model, the break luminosity is mainly determined by the cutoff luminosity of galaxies ($L_*$).
The indices of the double power-laws of quasar LFs are jointly determined by
$\alpha$ and the merger rates. 
In modeling quasar LFs, the free parameters ($\lambda$, $\epsilon_0$ and $t_{\rm dc,0}$)
degenerate weakly. However, their typical values can be quantified directly
from observations. 
We fix $\lambda \sim 0.15$ (Shen et al. 2008), $t_{\rm dc,0}\sim 3.0\times 10^7{\rm yrs}$
(Martini \& Weinberg 2001), and $\epsilon_0 \sim 10^{-5}$ (Ferrarese 2002). 
We calculate a grid of quasar LFs in a range of $\lambda$, $t_{\rm dc,0}$ and $\epsilon_0$ 
around their typical values and then obtain the fitting values accordingly.

The best fittings with minimal $\chi^2$ are given in Table 2 and shown in Figure 1. We find 
that the global features of quasar LFs can be well produced by Equation (13). The results 
modeled by galaxy LFs from Ilbert et al. (2005) and Dahlen et al. (2005) are 
in reasonable agreement. 
To match the observational data, we find that $q$ must be a function of redshift 
for both sets of galaxy LFs. 
We get a dependence of the mass ratios $q$ with respect to redshift as shown in Figure 2(a). It
is found that $q$ exhibits a rapid decline with redshift. 
A more explicit parameter $f_{\rm maj}$ clearly
shows significant evolution with redshift in Figure 2(b): it decreases from
$f_{\rm maj}\sim0.2$ at $z\sim2$ to $f_{\rm maj}\rightarrow0$ at $z\sim0$. This suggests a merger 
history with a trend from major to minor since $z\sim 2$.
Major merger rates measured from the Hubble Ultra Deep Field get a peak at 
{$z\sim 1-2$} (Ryan et al. 2008), 
further supporting the evolution of $f_{\rm maj}$ obtained here. 

The duty cycle time is obtained from Equation (14),  and the results are 
illustrated in Figure 3(a). We find that $t_{\rm dc}$ is strongly dependent on redshift, but insensitive
to the quasar luminosity, namely, the SMBH mass. The typical duty cycle time is
$\gtrsim10^7$years, indicating that the SMBHs
are undergoing many episodes (Martini 2004; Wang et al. 2006; 2008; 2009).
The lifetime of quasars is longer at higher redshift.
This should be due  to the increase of merger rates and the mounting importance of major
mergers with increasing redshift. Major mergers provide more abundant gas, and thus drive SMBH activities 
to live longer than minor mergers.
We also calculate the luminosity-averaged duty cycle time in Figure 3(b),
and simply fit its relation to redshift as
\begin{equation}\label{dutifit}
\bar{t}_{\rm dc}(z)=2\times 10^{6}\left(1+z\right)^{2.5}~{\rm yr}.
\end{equation}
The evolution of the duty cycle time should be independently examined by future
observational data, for example from the clustering measurements of quasars (for a review see Martini 2004). 
In addition, a more realistic model of the quasar lifetime
and light curve (Hopkins et al. 2005; Hopkins \& Hernquist 2009b) is worth considering in a future paper.

For the bin of $0.4\le z\le0.68$, we find
the modeled LFs under and overestimate the observed number of quasars at faint and bright 
tails, respectively. Bonoli et al. (2009) also find similar behaviors from semi-analytically modeling
quasar LFs coupled with numerical simulations. 
The over-estimate for the bright tail is because most of the galaxies
are gas-poor (dry mergers, see Lin et al. 2008) and some fraction of (low-level) quasar activities 
may be stochastical accretion-driven (Hopkins \& Hernquist 2006). 
This naturally drives an underestimate
of quasars for the faint tail regarding passive evolution of galaxies after $z\sim 1$
(Faber et al. 2007).
\subsection{Mergers and accretion history}
Major mergers are undoubtedly taking place in very luminous
submillimeter galaxies at $z\sim 2-3$ (Tacconi et al. 2006; 2008). 
Detailed studies from Spectroscopic Imaging survey in the
near-infrared with SINFONI (SINS) of high redshift galaxies ($z\sim1-3$)
show that about one third of the galaxies are 
rotation-dominated rings/disks undergoing star formations, another one third
are compact and velocity dispersion-dominated objects, and the remaining are clearly
interacting/merging systems (Genzel et al. 2008; F\"orster Schreiber et al. 2009).
It appears that rapid and continuous gas accretion via "cold
flows" and/or minor mergers are playing an important role in driving star formation
and the assembly of star forming galaxies (Genzel et al. 2008). Based on morphologies of X-ray
selected AGNs with redshifts $0.5\le z\le 1.3$, Georgakakis
et al. (2009) suggested that AGNs in disk galaxies are likely fueled by minor mergers.
More interestingly, Robaina et al. (2009) reported that only less than $10\%$ of star
formation in  massive galaxies for $z\sim 0.6$ is triggered by major interaction.
These pieces of independent evidence are consistent with the change of the
major merger fraction as shown in Figure 2{\em b}.

The importance of minor mergers in triggering SMBH activities has been realized by many authors
(e.g., De Robertis et al. 1998; Hernquist \& Mihos 1995; Taniguchi 1999; Kendall et al. 2003; 
Hopkins \& Hernquist 2009a). 
The dynamical studies of minor mergers find that the 
merging satellites retain memory of the initial random orientations when arriving at the
nucleus of the primary galaxy (Kendall et al. 2003). We therefore may infer that
during a succession of minor mergers, gas fueling to the central SMBH will carry 
on in a manner with stochastical orientations relative to the host galactic plane. 
This affords a natural explanation of random accretion proposed by 
King \& Pringle (2006; 2007), as one fueling scenario
to AGNs. There are several independent indications for minor mergers as a triggering mechanism 
of random accretion.

First, radio jets in a sample of nearby Seyfert galaxies ($z\leq0.03$) 
have a random orientation with respect to their host galactic plane (Kinney et al. 2000).
Presumably, the jet direction aligns with the axis of the accretion disk, therefore the fueling gas
to the SMBHs could {\em not} be supplied from their host galaxies directly (see a brief
review of fueling mechanisms of SMBHs in Taniguchi 1999), but probably from minor mergers (King \& Pringle 2007).

Second, AGN types are independent of orientations of their host galaxies 
(Keel 1980; Dahari \& De Robertis 1988; Munoz Marin et al. 2007). 
Additionally, the orientations of extended [O {\sc iii}] emission in some nearby Seyfert galaxies have 
no correlation with the major axis of the host galaxies (Schmitt et al. 2003). 
This suggests that the dusty torus axis is not aligned with the host galaxy rotation axis. 

Third, W09 derived an $\eta$-equation to
describe cosmological evolution of radiative efficiency of accretion into SMBHs, based on the argument 
that the SMBH growth is dominated by baryon accretion (Soltan 1982). 
Applying the equation into existing survey data of quasars and galaxies, they report 
that the radiative efficiency strongly evolves with redshift, with values of 
$\eta\sim 0.3$ at $z\sim 2$ and $\eta\sim 0.03$ at $z\sim 0$. As described below,
episodic random accretion definitely leads to a decrease of spin for SMBHs.
Presumably, with the link between radiative efficiency and SMBH spin (Thorne 1974),
this trend clearly supports the cosmological evolution of mergers from major 
at high-$z$ to minor at low-$z$ found from our model of quasars LFs.

With the above comprehensive indications and the results obtained in our model of quasar LFs, 
it quite appears that minor mergers may mainly be driving SMBH activities after $z\sim 2.0$, 
and random accretion driven by minor mergers supplies shining power of the quasars. 

\subsection{Spin evolution}
During the active phases of a quasar, the SMBH accretes a significant amount of
gas to buildup its mass (the Soltan's argument; see also Yu \& Tremaine 2002; Volonteri et al. 2003).
The previous studies using hierarchical merger tree found that mass accretion also plays a dominant 
role in determining spin evolution of black holes compared with coalescence (Volonteri et al. 2005; Berti \& Volonteri 2008). However, the detailed spin distribution of black holes
depends on the specific accretion scenarios: prolonged or random accretion. In prolonged accretion, the 
material keeps constant sign of angular momentum. In this case, an initially nonrotating
black hole that accretes about one-third of its mass will end up very fast spinning (Thorne 1974). 
By contrast, random accretion rapidly spins down black holes (Berti \& Volonteri 2008).
On the other hand, in a major coalescence some orbital angular momentum of a black hole binary 
may be converted to the spin of the newly formed black hole. 
This most likely yields a rapidly rotating black hole.

As mentioned above, random accretion can be naturally interpreted through stochastically-oriented 
minor mergers (Kendall et al. 2003). We have shown that minor mergers become important 
in driving SMBH activities toward low redshifts. Then we can estimate spin evolution of SMBHs
due to random accretion induced by minor mergers. 
For an approximation, we apply the equation $a=a_0(M_{\bullet,0}/\mbh)^{2.4}$ (Equation (26) in 
Hughes \& Blandford 2003, for random captures of minor black holes originally). 
Here $a=Jc/G\mbh^2$ is the spin parameter of a black hole with a mass $\mbh$ and 
angular momentum $J$, $c$ is the speed of light, $a_0$ and $M_{\bullet,0}$ are the initial 
spin and mass of the hole, respectively.
Under the assumption that each merger triggers a quasar activity, the number of merger events 
can be estimated through the ratio of the duty cycle time to the average lifetime of a single activity
\begin{equation}
n=\frac{\bar{t}_{\rm dc}}{\Delta t_{\rm Q}},
\end{equation}
where $\Delta t_{\rm Q}$ is the average lifetime of a single activity, 
$\bar{t}_{\rm dc}$ is given by Equation (19).
The lifetime of a single activity  can be inferred 
from the proximity effect, $\Delta t_{\rm Q}\sim$1 Myr (Kirkman \& Tytler 2008). 
For $z\approx2$, we  apply
$\bar{t}_{\rm dc}\approx3\times10^7$yr and then obtain the order of magnitude estimate $n\sim10$. 
To satisfy the ratio of SMBH mass and halo mass in Equation (6), the mass growth of SMBHs due to accretion 
during each merger is (see also Volonteri et al. 2003)
\begin{equation}
\Delta m=\frac{\gamma\epsilon}{3}\Delta\mh=\frac{\gamma}{3}\langle \xi\rangle\mbh,
\end{equation}
where $\langle \xi\rangle$ is the average mass ratio of two merging galaxies/halos.
The total mass growth during a Hubble time is thus
\begin{equation}
\frac{\Delta M_{\bullet}}{M_{\bullet}}=\frac{n\Delta m}{\mbh}=\frac{\gamma}{3}n\langle \xi\rangle\sim2,
\end{equation} 
where $\gamma=4.76$ and we set $\langle \xi\rangle\sim0.1$ for minor mergers.
Consequently, the spin of the black hole drops off by a factor of $3^{2.4}\sim 10$ from $z\approx2$ to $z\approx0$.
This is generally consistent with the results found by Berti \& Volonteri (2008) that random accretion 
predicts the spin distribution with $a\lesssim0.1$.

We may suggest a picture that at high redshift $z\sim 2$, the newly 
formed SMBHs from major mergers acquire a rapid spin
due to the orbital angular momentum of the merging holes, and subsequently random accretion 
led by minor mergers may rapidly spin them down.
It is interesting to note that King et al. (2008) proposed a scenario that 
self-gravity of a accretion disk can give rise to a long series of episodic activities 
during a major merger event. We neglect this effect in our model. 
A future work combining spin evolution with semi-analytical theory
will provide new insights into the coevolution among SMBHs, galaxies and dark matter halos.

\begin{deluxetable*}{lcccccc} 
\tabletypesize{\footnotesize}
\tablecolumns{7} 
\tablewidth{0.8\textwidth} 
\tablecaption{The relationship $\log (M_\bullet/M_\odot)=a+b M_R$
from published literature}
\tablehead{ 
\colhead{Reference} & \colhead{Sample\tablenotemark{a}}&
\colhead{Band\tablenotemark{b}} & 
\colhead{$a$}&\colhead{$b$} &\colhead{Scatter}  &\colhead{Note} 
} 
\startdata 
MD02   & 72S+18E & $R$ & $-2.96\pm0.48$  &$-0.50\pm0.02$  & 0.39    & \nodata\\
BFFF03 & 20E     & $R$ & $-3.00\pm1.35$  &$-0.50\pm0.06$  & 0.39    &\nodata\\
MH03   &12E+10S  & $B$ & $-2.54\pm1.10$  &$-0.50\pm0.05$  & 0.48    &\nodata\\
EGC04\tablenotemark{c}  &8E+5S    & $R$ & $+2.70$         &$-0.25$         &0.35  &\nodata\\
G07    &12E+10S  & $R$ & $+1.40\pm0.33$  &$-0.38\pm0.04$  & 0.33    & sample bias?\\
L07    &100E+55S+64B     & $V$ & $-3.70\pm0.77$  &$-0.55\pm0.03$  &\nodata  &\nodata \\
K08    & 45S     & $R$ & $-2.87\pm0.04$  &$-0.50$         & 0.41    & fix $\beta$\\
BPPV09 & 19E     & $V$ & $-1.67\pm1.58$  &$-0.47\pm0.08$  & 0.64    &fitting routine? \\
BPPV09 & 26A     & $V$ & $+1.22\pm0.75$  &$-0.32\pm0.04$  & 0.44    &fitting routine?
\enddata
\tablenotetext{a}{``E'' is for elliptical galaxies, ``S'' is for either S0 galaxies or Spiral galaxies,
``A'' is for AGNs, and ``B'' is for bright cluster galaxies.}
\tablenotetext{b}{$B-R=1.52$ and $V-R=0.61$ from Fukugita et al. (1995) are used to convert $B$- 
and $V$-band magnitude to $R$-band magnitude. The standard relation $\log L_V/L_\odot=0.4(-M_V+4.83)$ is
employed to calculate $M_V$ from $L_V$ (Bentz et al. 2009) and $M_{B,\odot}=5.47$ is used (Cox 2000).} 
\tablenotetext{c}{We artificially extract the data from the Figure (1.3) in EGC04.}
\tablerefs{MD02: McLure \& Dunlop (2002); BFFF03: Bettoni et al. (2003); MH03: Marconi \& Hunt (2003);
EGC04: Erwin et al. (2004);
G07: Graham 2007; L07: Lauer et al. (2007); K08: Kim et al. (2008); BPPV09: Bentz et al. (2009).}
\end{deluxetable*}

\section{Conclusions}
With the assumption that black holes are triggered through galaxy mergers, 
we successfully reproduce the observed quasar LFs by manipulating the galaxy LFs 
and merger rates from semi-analytic theory from $z\sim 0$ to $z\sim 2.0$.
This supports that mergers are driving black hole activities with many episodes.
We find that minor mergers become important in triggering quasar activities towards
low redshifts. As a natural consequence of minor mergers, the random accretion may 
drive growth and spin evolution of SMBHs at low redshift. The estimated spin declines 
rapidly with decreasing redshift, in agreement with the evolutionary trend of 
radiative efficiency for SMBHs.

\acknowledgements{We thank the anonymous referee for the helpful comments and suggestions that greatly 
improved the paper. YRL acknowledges H.-B. Yuan  and W.-W. Zuo for reading the 
manuscript and I. A. McNabb  for the help in improving the English presentation.
We also appreciate the stimulating discussions among the members
of IHEP AGN group. The research is supported by NSFC-10733010 and
10821061, CAS-KJCX2-YW-T03 and 973 project (2009CB824800). }
\section*{Appendix: Influences of $M_\bullet$-$L_{\rm bul}$ relationship on quasar LFs}

The relationship between SMBH mass and bulge luminosity ($M_\bullet$-$L_{\rm bul}$)
is an important ingredient in modeling quasar LFs from the galaxy LFs. 
Many previous works have focused on this relationship (see Table 3 for a brief summary). 
Unfortunately, not all of them reach the consistent conclusions.
At a first glance, the quiescent galaxies have a steeper slope of $M_\bullet$-$L_{\rm bul}$ than 
the AGNs. This may be due to neglecting the effect of radiation pressure, which causes underestimations 
of black hole mass in AGNs as proposed by Marconi et al. (2008, see, however, Netzer 2009). 
Meanwhile, Graham (2007) recently 
made an attempt to compensate for the analysis techniques employed by previous groups
(McLure \& Dunlop 2002; Marconi \& Hunt 2003; Erwin et al. 2004). 
They took into account a number of issues,
e.g. the removal of a dependency on the Hubble constant and a correction for the dust attenuation 
in bulges and disk galaxies. After careful revisions and adjustments, they found the differences 
among previous groups could be eliminated and they
reproduce a somewhat shallower slope. However, they did not exclude the potential sample bias, which 
may lead to a similar relation as McLure \& Dunlop (2002) [see Equation (A7) in Graham 2007].

To sum up, there appear to be many unclear aspects in the $M_\bullet$-$L_{\rm bul}$ relationship 
that deserve further examinations with improved observational data. We adopt McLure \& Dunlop's 
relation in the present work. 
To show the influences of $M_\bullet$-$L_{\rm bul}$ relationship on the quasar LFs, we simply 
perform an approximate estimate. For $\Delta \mh/\mh\ll1$, we have the merger rate (Lacey \& Cole 1993)
\begin{equation}
\dot{\cal R}\propto \left(\frac{\Delta\mh}{\mh}\right)^{-3/2}.
\end{equation}
Using the $\mbh-L_{\rm bul}$ relation and the mass ratio between the black hole and the dark matter halo, 
we have  
\begin{equation}
 \mh=\mbh/\epsilon\propto10^{b M_R}/\epsilon,
\end{equation}
and the galaxy LFs in $R$-band
\begin{eqnarray}\nonumber
 \Phi(M_R)&&\propto10^{0.4(1+\alpha)(M_R^*-M_R)}\exp\left[-10^{0.4(M_R^*-M_R)}\right]\\
          &&\propto\left(\frac{\mh^*}{\mh}\right)^{0.4(1+\alpha)/b}
             \exp\left[-\left(\frac{\mh^*}{\mh}\right)^{0.4/b}\right],
\end{eqnarray}
where $\mh^*=10^{a+b M_R^*}/\epsilon$.
Combining the above equations into Equation (13) and performing some mathematical expansions for 
$\Delta \mh/\mh\ll1$, we have the quasar LFs approximately as
\begin{eqnarray}\nonumber
\Psi&&\propto\int_0^{q\mh}\left(\frac{\Delta\mh}{\mh}\right)^{-1/2}
           \left(\frac{\mh^*/\mh}{1-\Delta\mh/\mh}\right)^{0.4(1+\alpha)/b}\\
    &&~~~\times\nonumber
           \exp\left[-\left(\frac{\mh^*/\mh}{1-\Delta\mh/\mh}\right)^{0.4/b}\right]\rd\Delta\mh\\\nonumber
    &&\propto\left(\frac{\mh^*}{\mh}\right)^{0.4(1+\alpha)/b}
      \exp\left[-\left(\frac{\mh^*}{\mh}\right)^{0.4/b}\right]\\
   &&\propto\left(\frac{L^*}{L}\right)^{0.4(1+\alpha)/b}
      \exp\left[-\left(\frac{L^*}{L}\right)^{0.4/b}\right].
\end{eqnarray}
We can find that the break luminosity $L^*$  of the quasar LFs is dependent on the cutoff luminosity $M_R^*$
of the galaxy LFs, and the slope $0.4(1+\alpha)/b$ is relative to the slope of the $\mbh-L_{\rm bul}$ relation.

\end{document}